\documentclass[preprints,entropy,review,submit,pdftex,moreauthors]{Definitions/mdpi}

\renewcommand{\linenumbers}{}

\firstpage{1} 
\pubvolume{1}
\issuenum{1}
\articlenumber{0}
\pubyear{2024}
\copyrightyear{2024}
\datereceived{ } 
\daterevised{ } 
\dateaccepted{ } 
\datepublished{ } 
\hreflink{https://doi.org/} 

\usepackage{bbm}
\usepackage{nicefrac}
\usepackage{xspace}

\DeclareMathOperator{\diag}{\mathrm{diag}}

\newcommand{\rep}[1]{\ensuremath\boldsymbol{#1}}
\newcommand{\crep}[1]{\ensuremath\bar{\boldsymbol{#1}}}
\newcommand{\Z}[1]{\ensuremath{\mathbbm{Z}_{#1}}} 

\newcommand{\SO}[1]{\ensuremath{\mathrm{SO}(#1)}}
\newcommand{\SU}[1]{\ensuremath{\mathrm{SU}(#1)}}

\newcommand{\e}{\mathrm{e}}
\newcommand{\I}{\mathrm{i}}
\newcommand{\Id}{\mathbbm{1}}

\newcommand{\CP}{\ensuremath{\mathcal{CP}}\xspace}
\newcommand{\x}{\ensuremath{\times}}

\newcommand{\SL}[2]{\ensuremath{\mathrm{SL}(#1, #2)}}

\usepackage{xcolor}

\Title{Flavor's Delight}

\TitleCitation{Flavor's Delight}

\Author{Hans Peter Nilles $^{1}$ and Sa\'ul Ramos-S\'anchez $^{2,}$*\orcidB{}}

\AuthorNames{Hans Peter Nilles and Sa\'ul Ramos-S\'anchez}

\AuthorCitation{Nilles, H.P.; Ramos-S\'anchez, S.}

\address{%
$^{1}$ \quad Bethe Center for Theoretical Physics, Physikalisches Institut der Universit\"at Bonn, Nussallee 12, 53115 Bonn, Germany; nilles@th.physik.uni-bonn.de\\
$^{2}$ \quad Instituto de F\'isica, Universidad Nacional Aut\'onoma de M\'exico, Cd.~de M\'exico C.P.~04510, M\'exico; ramos@fisica.unam.mx}

\corres{Correspondence: ramos@fisica.unam.mx}

\abstract{
Discrete flavor symmetries provide a promising approach to
understand the flavor sector of the standard model of particle
physics. Top-down (TD) explanations from string theory reveal two
different types of such flavor symmetries: traditional and
modular flavor symmetries that combine to the eclectic flavor
group. There have been many bottom-up (BU) constructions to fit
experimental data within this scheme. We compare TD and
BU constructions to identify the most promising groups and try
to give a unified description. Although there is some progress
in joining BU and TD approaches, we point out some gaps that
have to be closed with future model building.
}


\begin{document}

\section{Introduction}

The problem of flavor, the description of masses and mixing
angles of quarks and leptons, remains one of the most
important questions in elementary particle physics.
A major approach to solve this problem is based on non-Abelian
(discrete) flavor symmetries. In attempts to fit presently
available data, many different symmetries and representations
of flavor groups have been suggested and analysed. A comprehensive
summary of these BU attempts can be found in the
reviews~\cite{Feruglio:2019ybq,Kobayashi:2023zzc,Chauhan:2023faf,Ding:2023htn}. 
In his book~\cite{Frampton:2020kki} with Jihn E. Kim
entitled ``History of Particle Theory'', Paul Frampton (p.172) mentions
his preferred  flavor group $T'$, the binary tetrahedral group.
This choice is motivated through his early work on flavor
symmetries: see ref.~\cite{Frampton:2008bz} and references therein.

Most attempts in the BU approach focus on the lepton sector to
obtain solutions close to neutrino tribimaximal mixing~\cite{Harrison:2002er}.
Prominent examples have been $A_{4}$, $S_{4}$, $\Delta(27)$, $\Delta(54)$,
$\Sigma(81)$, $Q(24)$ among many others~\cite{Ma:2007hg}. While they lead to acceptable
solutions in the lepton sector, applications to the quark sector
have been less frequent and usually less successful. Still, as there
are many viable models it is difficult to draw a definite
conclusion about the correct choice.

It seems that we need additional ingredients to select models
from a more theoretical point of view. Such TD
considerations draw their motivation from string theory
model building, in particular orbifold compactifications of the
heterotic string~\cite{Bailin:1999nk,Nilles:2008gq,Ramos-Sanchez:2008nwx,Vaudrevange:2008sm,Nilles:2014owa,Ramos-Sanchez:2024keh}. 
Early work~\cite{Kobayashi:2006wq} on the \Z3 orbifold
revealed the discrete flavor group $\Delta(54)$ with irreducible
triplet representations to describe the three  families of
quarks and leptons. Even more earlier work,
analyzing duality symmetries in string theory~\cite{Lauer:1989ax,Lerche:1989cs,Chun:1989se,Lauer:1990tm},
provided an example of the discrete (modular) group $T'$.
From this point of view the predictions of the
\Z3 orbifold   lead to the discrete groups
$\Delta(54)$ and $T'$.

Fortunately, these groups allow many connections to models of the
BU approach, were in fact $T'$ and $\Delta(54)$ as well as
their ``little sisters'' $A_{4}$ and $\Delta(27)$ have
played a major role\footnote
{For an encyclopedia of discrete groups  and technical
details, we refer to ref.~\cite{Ishimori:2008uc}.}.
In the following we want to analyze these specific constructions
in detail. In section~\ref{sec:Tgroup} we start with the tetrahedral group $T$
(isomorphic to the group $A_4$ of even permutations of 4 objects)
that played a major role in the discussion of neutrino
tribimaximal mixing. We continue with its double cover
$T'$ and potential applications to flavor physics.
Section~\ref{sec:Delta27} introduces the motivation for the use of the group
$\Delta(27)$ for leptonic mixing. It has 27 elements and is
a discrete subgroup of \SU3. It is also a subgroup of
$\Delta(54)$ that appeared in early discussions of flavor groups in string theory
constructions~\cite{Kobayashi:2006wq}. Section~\ref{sec:TopDown} is devoted to TD
considerations of flavor symmetries from string theory model
building. There, we shall also introduce the concept of discrete
modular symmetries that were discovered from an analysis of dualities
in string theory~\cite{Lauer:1989ax,Lerche:1989cs,Chun:1989se,Lauer:1990tm}.
The application of modular symmetries to flavor physics was
pioneered in the BU approach by Feruglio~\cite{Feruglio:2017spp}
for the  example of the discrete modular group $A_{4}$.
We argue that the TD approach favors  instead the modular flavor group $T'$,
the double cover of $A_{4}$.
Section~\ref{sec:eclectic} introduces the concept of the eclectic flavor
group~\cite{Nilles:2020nnc,Nilles:2020kgo} that appears as a prediction in the
string theory framework. It combines the traditional flavor
symmetries (here $\Delta(54)$) with the discrete modular flavor symmetries
(here $T'$). In section~\ref{sec:TD-BU} we shall try to make contact between the BU
and TD approaches.
Section~\ref{sec:outlook} will give an outlook on strategies for further model
building. The appendices will give technical details of the properties of
$A_{4}$, $T'$, $\Delta(27)$ and $\Delta(54)$.

\section{The tetrahedral group and its double cover}
\label{sec:Tgroup}

The symmetry group $T$ of the tetrahedron is one of
the smallest non-Abelian
discrete groups and found early applications in
particle physics~\cite{Wyler:1979fe, Branco:1980ax}. It
has 12 elements and is isomorphic to $A_{4}$ the
group of even permutations of
four elements. There are three singlets $(\rep1, \rep1', \rep1'')$ and one
irreducible triplet representation. Detailed properties
of $T \cong A_{4}$ can be
found in Appendix~\ref{app:A4}. The presence of the triplet
representation makes it
attractive for flavor physics with three families of quarks
and leptons. It became particularly relevant for the discussion of (nearly)
tribimaximal mixing~\cite{Ma:2001dn,Altarelli:2005yp} in the lepton sector. 
An explicit discussion of this situation can be found
in the reviews~\cite{Feruglio:2019ybq,Ma:2007hg}. 
Tribimaximal mixing~\cite{Harrison:2002er} is characterized 
(up to phases) through the PMNS structure
\begin{equation*}
U_\mathrm{PMNS}~=~\begin{pmatrix}
 \sqrt{\tfrac23}   &\tfrac{1}{\sqrt3} & 0\\
 -\tfrac{1}{\sqrt6}&\tfrac{1}{\sqrt3} &-\tfrac{1}{\sqrt2}\\
 -\tfrac{1}{\sqrt6}&\tfrac{1}{\sqrt3} & \tfrac{1}{\sqrt2}
\end{pmatrix}
\end{equation*}
and includes a \Z2\x\Z2 symmetry acting (in the neutrino mass basis) as
$U = \diag(-1,-1,1)$ and $V = \diag(-1,1,-1)$. 
This symmetry is a subgroup of $S_{4}$,
the group of permutations of four
elements. Tribimaximal mixing, however, is not
exactly realized in nature as it
would imply that the (reactor) angle $\theta_{13}$
vanishes. The \Z2 transformation $V$ thus cannot be an exact symmetry.
This brings $A_{4}$ into the game,
a subgroup of $S_{4}$ that does not contain $V$.
It allows satisfactory fits for
the lepton sector as reviewed in ref.~\cite{Feruglio:2019ybq}.
These applications typically use the
triplet representation for the left-handed
lepton-\SU2-doublets $(\nu_i, \ell_i)$  and
the representations $(\rep1, \rep1'$ and $\rep1'')$
for the the \SU2
singlets of the standard model of particle physics (SM).
Various ``flavon'' fields have to be
considered for the spontaneous
breakdown of $A_{4}$ and this is subject to
explicit model building which we
shall not discuss here in detail. In any case,
$T \cong A_{4}$ is a very appealing
discrete flavor symmetry for the description of the lepton sector.

A look at the quark sector reveals a completely
different picture: there all
mixing angles are small and a fit similar to the lepton
sector does not seem to
work. One particular property of the quark sector is
the fact that the top-quark
is much heavier than the other quarks. This seems
to indicate a special role of
the third family, somewhat sequestered from the
other two families. It could therefore
imply that for quarks the third-family is a singlet
under the discrete flavor
group. Such a situation can be well described
in the framework of $T'$,
the double cover of $T \cong A_{4}$. This group
has 24 elements with representations
$\rep1, \rep1', \rep1'', \rep3$ (as $A_{4} \cong T$)
and in addition doublet
representations $\rep2, \rep2', \rep2''$
(details of properties of
$T'$ can be found in Appendix~\ref{app:T'}).
This double-cover is similar to the
double-cover \SU2 of \SO3 when
describing angular momentum. In fact, $T$ is
a subgroup of \SO3, and $T'$ a subgroup of \SU2. This implies that
the dynamics and constraints associated with $T$ can equally
stem from the larger group $T'$
(in analogy to the fact that one can also describe integer spin with \SU2), 
while the doublet representations of $T'$ allow for more 
options~\cite{Frampton:1994rk,Frampton:2007et}.

This fact has been used in ref.~\cite{Feruglio:2007uu,Carr:2007qw} 
to obtain a simultaneous description of
both, the lepton- and the quark-sector in the
framework of $T'$~\cite{Frampton:2008bz}. The
lepton sector remains the same as in the $A_{4}$
case while in the quark sector
we do not use the irreducible triplet representation,
but the representation $\rep1 \oplus \rep2'$, to single out the third family. 
This seems to be a nice explanation
of the difference of the quark and lepton sectors
within the flavor group $T'$. As Paul Frampton says in his
book with Jihn E.\ Kim~\cite{Frampton:2020kki}
(page 172) ``Clearly, it is better simultaneously
to fit both the quark- and
lepton-mixing matrices. This is possible using,
for example, the binary tetrahedral group $T'$''.
There are, of course, many other
attempts based on larger groups and
representations, but $T'$ remains a very
attractive option.

\section{\boldmath Towards larger groups: $\Delta(27)$ and $\Delta(54)$\unboldmath}
\label{sec:Delta27}

Although small groups such as $A_4$ and $T'$
already lead to satisfactory fits, there are a lot
of new parameters and ambiguities in explicit model
building and it is not evident whether this really gives
the ultimate answer. In fact, there have been many
more attempts with different groups and different
representations as can be seen in the reviews
ref.~\cite{Feruglio:2019ybq,Kobayashi:2023zzc,Chauhan:2023faf,Ding:2023htn}. 
Another attractive small group is $\Delta(27)$. It has 27 elements, 9 one-dimensional
representations as well as a triplet $\rep3$ and an
anti-triplet $\crep3$ representation. Technical
details of the group are given in Appendix~\ref{app:delta27}.
This is still a small group and it is attractive in
particular because of the $\rep3$ and $\crep3$
representations, which are well suited for flavor model building with
three families of quarks and leptons. As shown
in the appendix, it can be constructed as a semi-direct
product of \Z3\x\Z3 and \Z3 and it is a subgroup of \SU3.

Early applications can be found in ref.~\cite{Branco:1983tn,deMedeirosVarzielas:2006fc,Ma:2006ip,Ma:2007wu,Luhn:2007uq}
which exploit the presence of the $\rep3$ and $\crep3$
representations. For more recent work
and a detailed list of references, we refer to
ref.~\cite{deMedeirosVarzielas:2017sdv,CarcamoHernandez:2024vcr}.
As in the case of $A_4$,
also $\Delta(27)$ is well suited
to accommodate near tribimaximal mixing.
Again (as for $A_4$), the
\Z2\x\Z2 group of tribimaximal mixing is
not a subgroup of $\Delta(27)$, but it appears
approximately for specific alignments of vacuum
expectation values of flavon fields that appear
naturally within $\Delta(27)$.

$\Delta(27)$ is the ``little sister'' and subgroup
of $\Delta(54)$. This group has 54 elements, two
singlet, four doublet and two pairs of triplet and anti-triplet
($\rep3\oplus\crep3$) representations.
Properties of $\Delta(54)$ are collected in
Appendix~\ref{app:delta54}. It is already quite a large group,
somewhat unfamiliar to the BU flavor-community
and found less applications than $\Delta(27)$.
It became popular because of its appearance in
string theory~\cite{Kobayashi:2006wq}, which we shall discuss
in section~\ref{sec:TopDown} in detail. Explicit BU model building
with $\Delta(54)$ was pioneered in ref.~\cite{Ishimori:2008uc}.

\section{Top-Down considerations: A taste of flavor from string theory}
\label{sec:TopDown}

In the BU approach there are many successful models based
on various groups and representations~\cite{Feruglio:2019ybq,Kobayashi:2023zzc,Chauhan:2023faf,Ding:2023htn},
too many to single out a ``best'' option. Such an answer might come
from theoretical considerations and top-down model building.
An attractive framework is given by string theory. Here we
shall concentrate on orbifold compactifications of heterotic
string theory that provide many realistic models with
gauge group \SU3\x\SU2\x$\mathrm{U}(1)$ and three families
of quarks and leptons~\cite{Buchmuller:2005jr,Lebedev:2006kn,Lebedev:2007hv,Lebedev:2008un,Nilles:2008gq,Nilles:2014owa,Olguin-Trejo:2018wpw,Perez-Martinez:2021zjj}.

In these theories, discrete flavor symmetries arise as a
result of the geometry of extra dimensions and the geography of
fields localized in compact space. Strings are extended
objects and this reflects itself in generalized aspects of
geometry that include the winding modes of strings. A full
classification of flavor symmetries of orbifold
compactifications of the heterotic string is given by the
outer automorphisms~\cite{Baur:2019kwi,Baur:2019iai} of the
Narain space group~\cite{Narain:1985jj,Narain:1986am,Narain:1986qm,GrootNibbelink:2017usl}.
Here we shall not be able to give a full derivation of
this fact, but only provide a glimpse of the general
TD formalism and illustrate the results in simple examples
based on a $D=2$-dimensional torus and its orbifold.

In general, a string in $D$ dimensions has $D$ right-moving and $D$ left-moving degrees of freedom,
encoded in $Y=(y_\mathrm{R}, y_\mathrm{L})$. Compactifying the theory on a $D$-dimensional 
torus demands that the $2D$ degrees of freedom be subject to the toroidal boundary conditions
$$
Y ~=~
\begin{pmatrix}
      y_\mathrm{R} \\      
      y_\mathrm{L} 
\end{pmatrix}
~\sim~ Y + E\, \hat N ~=~
\begin{pmatrix}
      y_\mathrm{R} \\      
      y_\mathrm{L} 
\end{pmatrix}
+ E
\begin{pmatrix}
      n \\      
      m
\end{pmatrix},
$$
where the winding and the Kaluza-Klein (momentum) quantum numbers of the string, $n,m\in\Z{}^D$, define a $2D$-dimensional
Narain lattice. $E$ denotes the so-called Narain vielbein and contains the moduli $M_i$ of the torus. In the
Narain formulation, we achieve a $D$-dimensional orbifold by imposing the identifications
$$
Y \sim
\Theta^k\, Y + E\, \hat N, \quad \text{with the \Z{K} orbifold twist}\quad \Theta =
\begin{pmatrix}
  \theta_\mathrm{R} & 0 \\     
  0                 & \theta_\mathrm{L} 
\end{pmatrix}\quad \text{satisfying} \quad \Theta^K = \Id_{2D}\,,$$
where $k=0,\ldots,K-1$ and the \SO{D} elements $\theta_\mathrm{L}, \theta_\mathrm{R}$ are
set to be equal to obtain a symmetric orbifold. Excluding roto-translations, the Narain space group can then be generated by
$$
{\rm the\ twist}\ \ (\Theta, 0)\ \ {\rm and \ shifts}\ \  (\Id, E_i) \ \ {\rm for} \ \ i = 1, \ldots, 2D\;.
$$
It turns out that flavor symmetries correspond to the (rotational and translational)
outer automorphisms of this Narain space group~\cite{Baur:2019kwi,Baur:2019iai},
which are transformations that map the group to itself but do not belong to the group.

Hence, it follows that the flavor symmetries
of string theory come in two classes:
\begin{itemize}
\item Those symmetries that map momentum- to momentum-
and winding- to winding-modes. These symmetries we call
traditional flavor symmetries. They are the same type as
those symmetries that would appear in a quantum field theory
of point particles. In the Narain formulation, these
can be understood as translational outer automorphisms of
the Narain space group.

\item Symmetries that exchange winding- and
momentum-modes. They have their origin in duality
transformations of string theory. We call them modular
flavor symmetries as (for the torus discussed here)
they are connected to the modular group \SL{2}{\Z{}}.
These arise from rotational outer automorphisms of
the Narain space group.
\end{itemize}

\subsection{Traditional flavor symmetries}

Here we concentrate on the two-dimensional cases
$\mathbbm{T}^2/\Z{K}$, $K=2,3,4,6$, that could be understood as the
fundamental building
blocks for the discussion of flavor symmetries. They
have been discussed in detail in ref.~\cite{Kobayashi:2006wq}.
Various groups can be obtained, prominently $D_8$ or
$\Delta(54)$. As an illustrative example we discuss
here the case $\mathbbm{T}^2/\Z{3}$ with group $\Delta(54)$ because it has the
nice property to provide irreducible triplet representations for three
families of quarks and leptons~\cite{Carballo-Perez:2016ooy,Olguin-Trejo:2018wpw}.

\begin{figure}
\centerline{\includegraphics[height=1.5in]{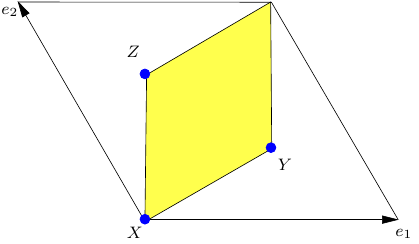}}
\caption{The $\mathbb{T}^2/\Z3$ orbifold
(yellow shaded region) with three fixed points $X,Y,Z$. Twisted states are localized
at theses fixed points.
Figure taken from ref.~\cite{Nilles:2021ouu}. 
\label{fig:Z3}}
\end{figure}

The \Z3 orbifold $\mathbbm{T}^2/\Z{3}$ is shown in figure~\ref{fig:Z3}.
Twisted fields are localized on the fixed points
$X, Y, Z$ of the orbifold. This geometry leads to
an $S_3$ symmetry from the interchange of the fixed
points. String theory selection rules provide an
additional \Z3\x\Z3 flavor symmetry as discussed
in ref.~\cite{Kobayashi:2006wq}. The full traditional flavor symmetry
is $\Delta(54)$, the multiplicative closure of these
groups. The twisted states on the fixed points
$X, Y, Z$ transform as (irreducible) triplets under
$\Delta(54)$ (details can be found in Appendix~\ref{app:delta54}).
$\Delta(54)$ has two independent triplet representations
$\rep3_1$ and $\rep3_2$. Both can be realized in string theory,
depending on the presence or absence of twisted
oscillator modes~\cite{Nilles:2020kgo}.
The untwisted states are in the trivial $\rep1$ representation
in the absence and $\rep1'$ in the presence of
oscillator modes. A nontrivial vacuum expectation value
of a field in the $\rep1'$ representation will break
$\Delta(54)$ to $\Delta(27)$. A discussion of the breakdown
pattern of $\Delta(54)$ can be found in ref.~\cite{Baur:2021bly}. 
Winding states transform as doublets under
$\Delta(54)$. They are typically heavy
and could play a prominent role in the discussion of
\CP-violation in string theory as they provide a mechanism
for baryogenesis through the decay of the heavy winding
modes~\cite{Nilles:2018wex}.

\subsection{Modular flavor symmetries}

They have their origin in duality transformations of string
theory. One example is $T$-duality that exchanges winding and
momentum modes. As a warm-up example consider a string on a
circle of radius $R$.

\begin{figure}[t!]
\centering
\includegraphics[height=3in]{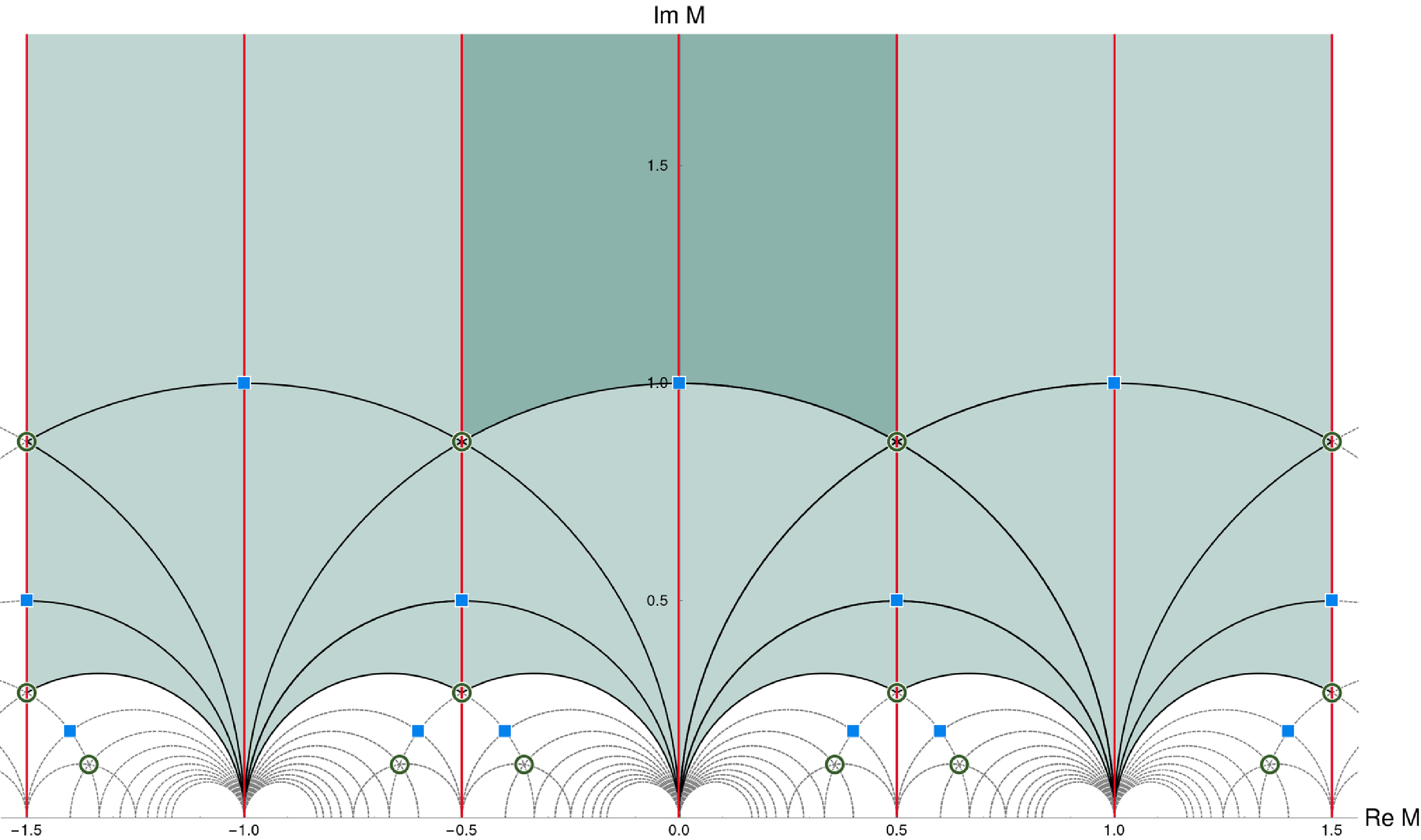}
\caption{Fundamental domain of $\SL{2}{\Z{}}$ (dark shaded) and 
of its subgroup $\Gamma(3)\cong\SL{2}{3\Z{}}$
(light shaded). 
Figure taken from ref.~\cite{Nilles:2021ouu}}. 
\label{fig:ModuliFundamentalDomain}
\end{figure}

The masses of momentum modes are governed by $1/R$, while winding
states become heavier as $R$ grows. On the other hand, T-duality
of string theory is defined by the transformations
$$
\text{winding modes} ~\longleftrightarrow~ \text{momentum modes}
\qquad \text{and}\qquad
R ~\longleftrightarrow~ \alpha'/R\;.
$$
Hence, T-duality maps a theory to its T-dual, which coincides at
the self-dual point $R^2 = \alpha' = 1/M_{\rm string}^2$,
where $1/\alpha'$ is the string tension. For a generic value of
the modulus $R$, T-duality exchanges light and heavy states, which
suggests that T-duality could be relevant to flavor physics.
Since string theory demands the compactifications of more than
one extra dimension, T-duality generalizes to large groups of
nontrivial transformations of the moduli of higher-dimensional tori. For instance, in $D=2$
the transformations on each of the (K\"ahler and complex structure) moduli
build the modular group $\SL{2}{\Z{}}^2$ of the $\mathbbm{T}^2$ torus.
The group \SL{2}{\Z{}} is generated by two elements
$$
\mathrm{S}\ \text{ and }\ \mathrm{T}, \qquad \text{such that}\qquad
\mathrm{S}^4 ~=~ \Id\,,\qquad \mathrm{S}^2\mathrm{T} ~=~ \mathrm{T}\mathrm{S}^2
\quad \text{and} \quad(\mathrm{S}\mathrm{T})^3 ~=~ \Id\;.
$$
For each modular group \SL{2}{\Z{}}, there exists an associated modulus $M$ that
transforms as
$$\mathrm{S}:\quad M ~\mapsto~ -{1\over M} \qquad \text{and} \qquad
  \mathrm{T}:\quad M ~\mapsto~ M+1\;.
$$
Further transformations include mirror symmetry (that exchanges K\"ahler and complex
structure moduli) as well as the \CP-like transformation
$$
\mathrm{U}:\quad M ~\mapsto~ -\overline{M}\,,
$$
where $\overline{M}$ denotes the complex conjugate of $M$.
String dualities give important constraints on the action
of the theory via the modular group \SL{2}{\Z{}} (or $\mathrm{GL}(2,\Z{})$
when including $\mathrm{U}$). A general \SL{2}{\Z{}} transformation
of the modulus is given by
$$
M ~\stackrel{\gamma}{\longmapsto}~ {{a\,M+b}\over{c\,M+d}},\qquad
\gamma ~=~\begin{pmatrix}a&b\\c&d\end{pmatrix}~\in~\SL{2}{\Z{}}
$$
with $\det\gamma=1$ and $a,b,c,d\in\Z{}$. The value of $M$ (originally in the
upper complex half plane) is then restricted to the fundamental domain, as shown in
(the dark shaded region of) figure~\ref{fig:ModuliFundamentalDomain}.
Matter fields $\phi$ turn out to transform as
$$
\phi ~\stackrel{\gamma}{\longmapsto}~ (c\,M+d)^k \rho(\gamma)\,\phi \qquad{\rm for}\qquad \gamma ~\in~ \mathrm{SL}(2,\Z{})\;,
$$
where $(c\,M+d)^k$ is known as automorphy factor, $k$ is a modular weight fixed
by the compactification properties~\cite{Ibanez:1992hc,Olguin-Trejo:2017zav},
and $\rho(\gamma)$ is a unitary representation of $\gamma$. Interestingly,
$(\rho(\mathrm{T}))^N = \Id$ even though $\mathrm{T}^N\neq\Id$, such that
$\rho(\gamma)$ generates a so-called finite modular group, as we shall shortly
discuss. Among others, the modular weights $k$ of the fields are important 
ingredients for flavor model building.

As in the one-dimensional case, duality maps one theory to its dual
and there remains the question whether such
transformations are relevant for the low-energy
effective action of the massless fields. This has been
discussed explicitly with the help of worldsheet
conformal field theory methods~\cite{Lauer:1989ax,Lerche:1989cs,Chun:1989se,Lauer:1990tm}.
It leads to field-dependent Yukawa couplings that
transform as modular forms $Y^{(n_Y)}(M)$ of a weights $n_Y>0$
$$
Y^{(n_Y)}(M) ~\stackrel{\gamma}{\longmapsto}~ (c\,M+d)^{n_Y} \rho_Y(\gamma)\,Y^{(n_Y)}(M)
\quad \text{for}\quad \gamma ~\in~ \SL{2}{\Z{}}\;,
$$
where, as for matter fields, $\rho_Y(\gamma)$ is also a unitary
representation of $\gamma$ in a finite modular group.
The description in terms of supergravity actions has been
given in ref.~\cite{Ferrara:1989bc}. From the transformation
properties of matter fields and Yukawa couplings, it becomes
clear that the action is subject to both invariance under the
finite modular group and conditions on the modular weights,
which are strongly restricted in the TD approach.

Let us illustrate the relevance to flavor physics in
the case of the \Z3 orbifold. We start with a
two-torus and its two moduli: K\"ahler modulus $M$ and
complex structure modulus $U$. On the orbifold
the $U$-modulus is frozen, such that the lengths of the lattice
vectors $e_1$ and $e_2$ are equal with an angle of
120 degrees (see figure~\ref{fig:Z3}). This also gives restrictions
on the modular transformations of the matter fields.
The coefficients $a,b,c,d\in\Z{}$ of the modular transformation
are defined only modulo 3; hence, instead of the full
modular group \SL{2}{\Z{}}, we have to deal with its
so-called principal congruence subgroup,\footnote{The principal
congruence subgroup of level $N$ is denoted by $\Gamma(N)$ and 
built by all $\gamma\in\SL{2}{\Z{}}$, such that $\gamma=\Id\mod N$.}
$\Gamma(3)\cong\SL{2}{3\Z{}}$. Clearly, $\Gamma(3)$ has still 
infinitely many elements, but it is a normal subgroup
of finite index in \SL{2}{\Z{}}. Hence, a finite
discrete modular group can be obtained by the quotient
$\SL{2}{\Z{}}/\Gamma(3)=\Gamma_3'$. An explicit discussion
is given in ref.~\cite{deAdelhartToorop:2011re}.
$\Gamma_3'$ is isomorphic to $T'$, the binary tetrahedral
group. It is the double cover of $\Gamma_3\cong A_4$, which
one would obtain starting with $\mathrm{PSL}(2,\Z{})$ instead of
\SL{2}{\Z{}}. In the first application of discrete modular
flavor symmetry Feruglio~\cite{Feruglio:2017spp} used the group
$\Gamma_3\cong A_4$ with its representations
$\rep1,\rep1',\rep1''$ and $\rep3$ to explain
tribimaximal mixing in the standard way. Complications
with flavon fields and many additional parameters
could be avoided as the modular flavor symmetry is
nonlinearly realized. This might lead to problems
with the control of additional free parameters in the
K\"ahler potential that has been taken into account~\cite{Chen:2019ewa}.
The modular flavor approach was picked up quickly
\cite{Kobayashi:2018vbk,Penedo:2018nmg,Ding:2019xna,Liu:2019khw,Liu:2020msy,Liu:2021gwa,Kobayashi:2023zzc,Ding:2023htn,Arriaga-Osante:2023wnu}
and led to many different BU constructions with
various groups, representations of modular weights.

Unfortunately, the TD approach is much more restrictive
and allows less freedom in model building. In our example
we obtain $T'$ and not $A_4$ (the double cover is
necessary to obtain chiral fermions in the string
construction). Moreover, the twisted states do not
transform as irreducible triplets of $T'$ but as
$\rep1\oplus\rep2'$ and the modular weights of the fields are
correlated with the $T'$ representation (thus
cannot be chosen freely as done in the BU framework).
Some details of $T'$ modular forms are given in
Appendix~\ref{app:TprimeForms}.

\section{Eclectic flavor groups}
\label{sec:eclectic}

So far we have seen that string theory predicts the presence of
both, the traditional flavor group ($\Delta(54)$ in our example)
and the modular flavor group ($T'$). You cannot have one
of them without the other. This should be taken into account in
flavor model building. The eclectic flavor group~\cite{Nilles:2020nnc}
is the multiplicative closure of $\Delta(54)$ and $T'$,
here $\Omega(1)=[648,533]$.\footnote{We have somewhat simplified
the discussion here. In full string theory with six compact extra
dimensions, we usually find additional $R$-symmetries that would
increase the eclectic flavor group, here the group
$\Omega(1)$ to $\Omega(2)=[1944,3448]$. A detailed discussion of these
subtleties can be found in ref.~\cite{Nilles:2020tdp,Nilles:2020gvu}.
}
Observe that this group has only 648 elements for the product
of groups with 54 and 24 elements, respectively. There is one
\Z2-like element contained in both $\Delta(54)$ and $T'$.
Incidentally, this is the same element that enhances $\Delta(27)$
to $\Delta(54)$. Thus $\Delta(27)$ and $\Delta(54)$, together
with $T'$, would lead to the same eclectic group~\cite{Nilles:2020nnc}.

\begin{figure}[h!]
\centering
\includegraphics[height=3.9in]{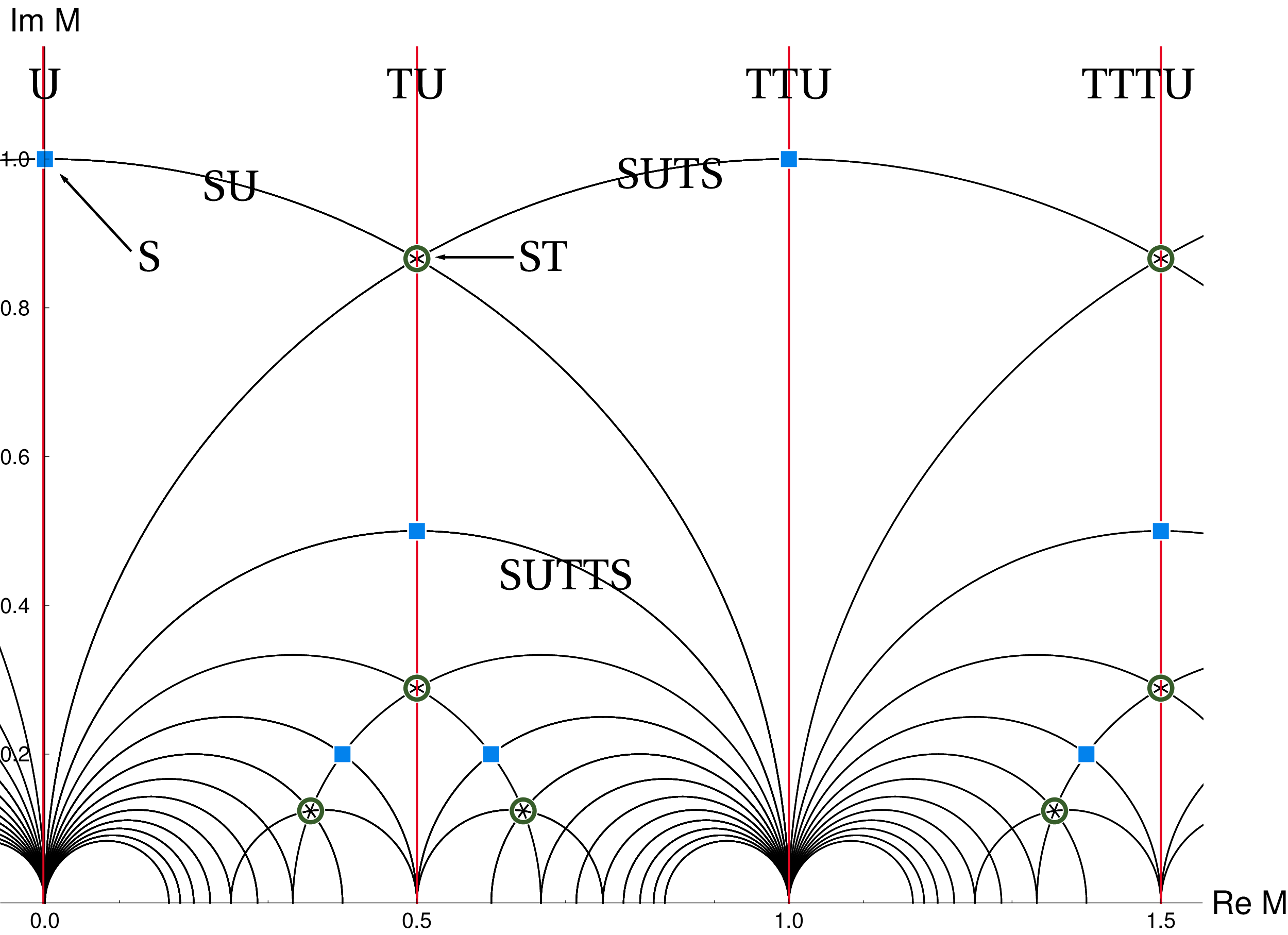}
\caption{Unbroken modular symmetries at special curves in moduli space, including
the \CP-like generator $\mathrm{U}$, which maps $M \mapsto -\overline{M}$. Figure adapted
from ref.~\cite{Baur:2019iai}.
\label{fig:fixedPoints}}
\end{figure}

The eclectic flavor group is nonlinearly realized.
Part of it appears ``spontaneously'' broken through the vacuum
expectation value of the modulus $M$. The modulus is confined
to the fundamental domain of $\Gamma(3)=\SL{2}{3\Z{}}$ as displayed
in figure~\ref{fig:ModuliFundamentalDomain}.
This area is reduced by a factor two if we include
the natural candidate for a \CP-symmetry that transforms
$M$ to $-\overline M$. The \CP-symmetry extends \SL{2}{\Z{}} to
$\mathrm{GL}(2,\Z{})$, $T'$ to $\mathrm{GL}(2,3)$ (a group with 48 elements)
and the eclectic group $\Omega(1)$ to a group with 1296
elements. The fundamental domain includes fixed points and
fixed lines with respect to the modular transformations
$\mathrm{S}$ and $\mathrm{T}$ as well as the \CP-transformation
$\mathrm{U}: M \to -\overline{M}$ as shown in figure~\ref{fig:fixedPoints}.

\begin{figure}[t!]
\centering
\includegraphics[height=4in]{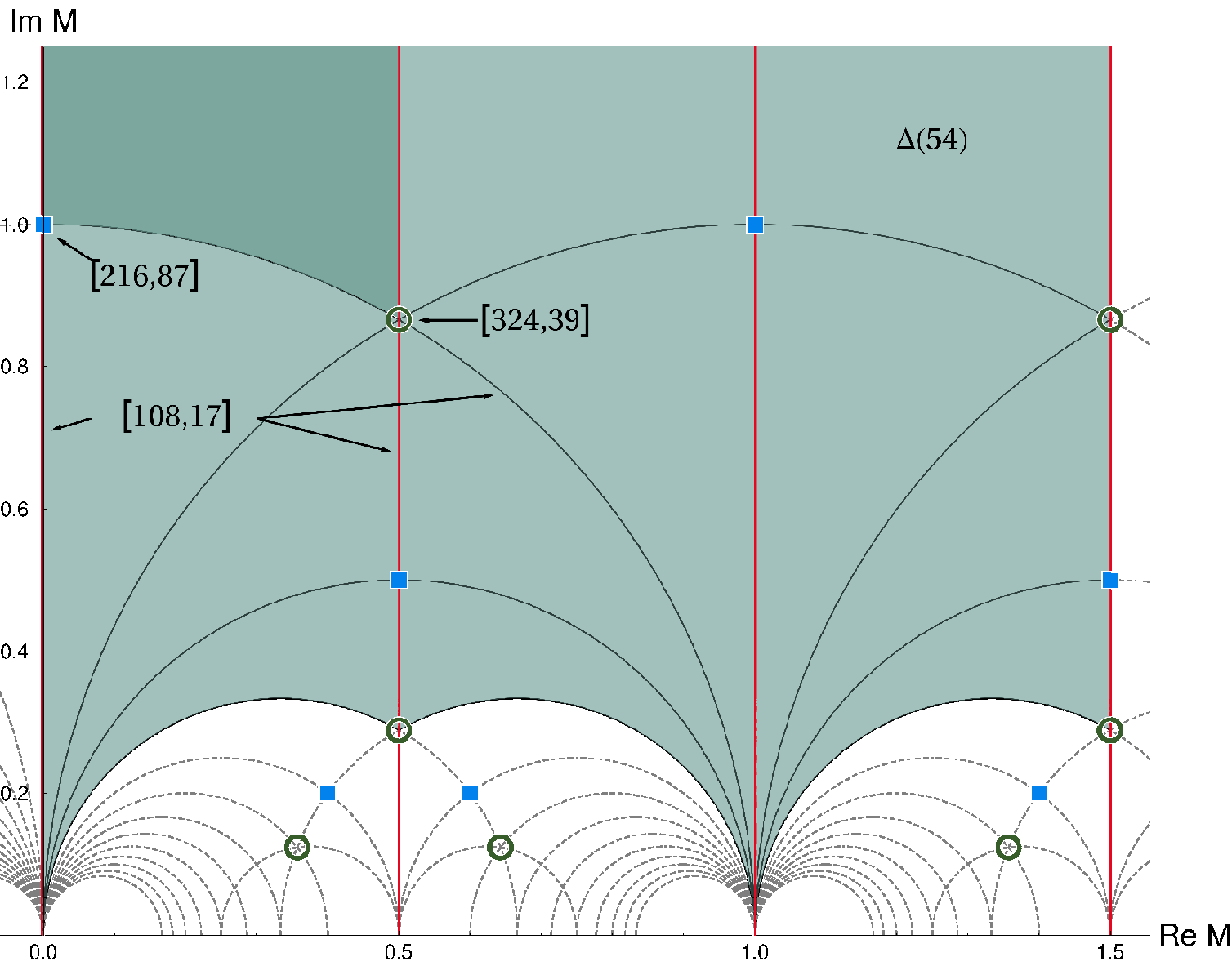}
\caption{Local flavor unification at special points and curves in moduli space. The traditional flavor
symmetry $\Delta(54)$, valid at generic points, is enhanced to two (different) groups with GAP Id [108,17]
at the vertical lines and semi-circles, including \CP-like transformations. At the intersections of curves,
the flavor symmetry is further enhanced to [216,87] and [324,39], also with \CP-like transformations.
Figure adapted from ref.~\cite{Baur:2019kwi}.
\label{fig:Z3Enhancements}}
\end{figure}

For generic points in moduli space the traditional
flavor symmetry $\Delta(54)$ is linearly realized. At the
fixed points and lines this symmetry is enhanced to larger
groups as illustrated in figure~\ref{fig:Z3Enhancements}.

We see that here the largest linearly realized group has 
324 elements with GAP Id~[324,39]. (We use the group notation of the
classification of GAP~\cite{GAP4}.) Thus only part of the
full eclectic flavor group with 1296 elements (including \CP) can be
linearly realized. The enhancement of the traditional
flavor symmetry at fixed loci (here points and lines)
in the fundamental domain exhibits the phenomenon called
``Local Flavor Unification''~\cite{Baur:2019kwi,Baur:2019iai}. 
The flavor symmetry is non-universal in moduli space and the spontaneous
breakdown of modular flavor symmetry can be understood
as a motion in moduli space. This has important
consequences for flavor model building. At the loci
of enhanced symmetry some of the masses and mixing
angles of quark- and lepton-sector might vanish. The
explanation of small parameters and hierarchies in
flavor physics can thus find an explanation if the modulus
is located close to the fixed points or
lines~\cite{Feruglio:2021dte,Baur:2021bly,Novichkov:2022wvg,Baur:2022hma,Feruglio:2022koo,Hoshiya:2022qvr,Petcov:2023vws,Abe:2023qmr,deMedeirosVarzielas:2023crv,Ding:2024xhz}.
The mechanism of moduli stabilization in string theory
could therefore provide the ingredients to understand
the mysteries of flavor~\cite{Kobayashi:2019xvz,Novichkov:2022wvg,Knapp-Perez:2023nty,Kobayashi:2023spx}.

\section{Top-down does not yet meet bottom-up}
\label{sec:TD-BU}

There have been many BU constructions, but only a few that
took into account TD considerations~\cite{Chen:2021prl,Ding:2023ynd,Li:2023dvm}. From the presently
available TD models, the groups $\Delta(54)$ for traditional
and $T'$ for modular flavor symmetry seem to be the
favourite choices. In fact, there is only one explicit model
that incorporates the SM
with gauge group \SU3\x\SU2\x$\mathrm{U}(1)$ and three
families of quarks and leptons~\cite{Baur:2022hma}. We certainly need more
work in the TD approach. Therefore, any conclusions about
the connection between the two approaches is necessarily
preliminary. Still, it is reassuring to see that the same
groups $\Delta(54)$ and $T'$ and their ``little
sisters'' $\Delta(27)$ and $A_4$ appear prominently in
BU constructions. One could therefore try to make contact
between the two approaches within this class of models.

Before we do that, we would like to stress some important
properties  of the TD approach that seem to be of more
general validity and thus should have an influence on
BU model building. The first of this is the prediction
of string theory for the simultaneous presence of both,
traditional and modular flavor symmetry that combine to
the eclectic flavor group. It is this eclectic group
which is relevant, not one of the others in isolation.
Up to now, many BU constructions only consider one of
them. Therefore, a direct contact between the two
approaches is very difficult at this point.

The TD approach is very restrictive. Apart from the
the limited type of groups that appear in the
TD constructions, there are also severe restrictions
on the choice of representations. Not everything is
possible. In the case of modular symmetry $T'$, for
example, the irreducible triplet representation does
not appear in the spectrum, while many BU constructions
exactly concentrate on this representation. Therefore,
the TD approach cannot make contact with models based
on modular $A_4$ flavor symmetry where these triplets
are generally used. For $T'$ we have the twisted fields in 
the $\rep1 \oplus \rep2'$ representation.
It seems to be more likely that irreducible triplet
representations are found within  the traditional
flavor group, as seen in the example with $\Delta(54)$.

A second severe restriction concerns the choice of
modular weights. In the TD approach we have
essentially no choice. Once we know the representations
of the eclectic group, the modular weights are fixed.
This is an important restriction as in the BU approach
the choice of modular weights is an important ingredient
of model building. With a careful choice of modular
weights one can create additional ``shaping symmetries''
which are important for the success of the fit to the
data. This is not possible in the TD approach. There
the role of such symmetries could, however, be found
in the traditional flavor symmetry.

As a result of these facts, there is presently still a 
crucial difference between the BU and TD approaches and a
direct comparison is not possible at this point.
We are still at the very early stage of such investigations.

\section{Outlook}
\label{sec:outlook}

Much more work in both approaches is needed to clarify the
situation. In the BU approach it would be desirable to
consider models that fulfill the restrictions coming from
TD. Traditional flavor symmetries and the eclectic
framework should be taken into account. A toolkit for
such a construction can be found in the consideration
of a modular group that fits into the outer automorphism
of the traditional flavor group, as explained in ref.~\cite{Nilles:2020nnc}.
A recent application of this connection for the
traditional flavor group
$\Delta(27)$ has been discussed in ref.~\cite{Ding:2023ynd,Li:2023dvm}.
Moreover, BU constructions should avoid the excessive
use of modular weights in model building. A strict
correlation between the representations and their
modular weights might be the right way to proceed.
Useful shaping symmetries might be found within the
traditional flavor symmetry instead.

The TD approach needs serious attempts for the
construction of more explicit models. In particular, it
would be useful to increase the number of explicit
string constructions that ressemble the SM
with gauge group \SU3\x\SU2\x$\mathrm{U}(1)$ and three families of
quarks and leptons. This is important, as in generic
string theory we might find huge classes of duality
symmetries that might not survive in models with the
properties of the SM.
Of course, the size and nature of these large symmetry groups
have to be explored. Modular invariance and its group
$\SL{2}{\Z{}}$ are closely related to torus compactifications,
that can be realized in orbifold compactifications and
more generally in Calabi-Yau compactifications with
elliptic fibrations. This can be described by the
basic building blocks $\mathbbm T^2/\Z{K}$ with $K=2,3,4,6$,
which have been studied so far~\cite{Nilles:2020gvu}.
Explicit string model building shows that such situations are
possible, but require particular constellations for the
Wilson lines needed for realistic model building.
Such Wilson lines and other background fields
might otherwise break modular symmetries in various ways~\cite{Bailin:1993ri,LopesCardoso:1994is,Love:1994ms}. 
In some orbifolds, only a subgroup of $\SL{2}{\Z{}}$ is unbroken
even without background fields~\cite{Bailin:1994ma}, which opens up
the possibility of finite modular flavor symmetries beyond $\Gamma'_K$
~\cite{Liu:2021gwa,Ding:2023ydy,Arriaga-Osante:2023wnu}. Yet
a more general discussion has
to go beyond $\SL{2}{\Z{}}$. A first step in the direction
is the consideration of the Siegel modular group~\cite{Baur:2020yjl,Nilles:2021glx,Ding:2024xhz}
or higher dimensional constructions~\cite{deMedeirosVarzielas:2019cyj,Nilles:2020tdp,Nilles:2020gvu,Kikuchi:2023awe}.
Many exciting developments seem to be in front of us.

\vspace{6pt} 

\funding{The work by SR-S was partly funded by UNAM-PAPIIT grant IN113223 and Marcos Moshinsky Foundation.}

\dataavailability{No new data were created or analyzed in this study. Data sharing is not applicable to this article.}

\acknowledgments{We acknowledge Alexander Baur, Mu-Chun Chen, Moritz Kade, Victoria Knapp-P\'erez,
                 Xiang-Gan Liu, Yesenia Olgu\'in-Trejo, Ricardo P\'erez-Mart\'inez, Mario Ramos-Hamud,
                 Michael Ratz, Andreas Trautner, Patrick Vaudrevange
                 for fruitful, interesting and pleasant collaborations.}

\conflictsofinterest{The authors declare no conflicts of interest.}

\abbreviations{Abbreviations}{
The following abbreviations are used in this manuscript:\\

\noindent
\begin{tabular}{@{}ll}
TD & Top-Down\\
BU & Bottom-Up\\
SM & Standard Model\\
PMNS & Pontecorvo-Maki-Nakagawa-Sakata
\end{tabular}
}

\appendixtitles{yes} 
\appendixstart
\appendix
\section[\appendixname~\thesection]{\boldmath The group $A_4$ and its double cover $T'$ \unboldmath}
\label{app:groupTheory}

\subsection[\appendixname~\thesubsection]{ $A_4$ }
\label{app:A4}

$A_4\cong(\Z2\x\Z2)\rtimes\Z3$ (GAP Id~[12,3]) is the alternating group of four elements and can also be understood as the rotational
symmetry group of a regular tetrahedron. It contains 12 elements. $A_4$ has the irreducible representations $\rep{r}\in\{\rep1,\rep1',\rep1'',\rep3\}$. 
With $\omega:=\e^{\nicefrac{2\pi\I}{3}}$, its character table reads
\begin{center}
\begin{tabular}{c|cccc}
  class          & $1C_1$ & $3C_2$       & $4C_3$       & $4C_3'$ \\
  representative & $\Id$    & $\mathrm{S}$ & $\mathrm{T}$ & $\mathrm{T}^2$     \\
  \hline
  $\rep1$   & $1$ & $1$  & $1$        & $1$ \\
  $\rep1'$  & $1$ & $1$  & $\omega$   & $\omega^2$ \\
  $\rep1''$ & $1$ & $1$  & $\omega^2$ & $\omega$ \\ 
  $\rep3$   & $3$ & $-1$ & $0$        & $0$
\end{tabular}
\end{center}
in terms of the generators $\mathrm{S}$ and $\mathrm{T}$. They satisfy
$\mathrm{S}^2 = \mathrm{T}^3 = (\mathrm{ST})^3 = \Id$ and their representations $\rho_{\rep{r}}$
can be expressed by
\begin{center}
\begin{tabular}{c|cccc}
  $\rep{r}$ & $\rep1$ & $\rep1'$ & $\rep1''$  & $\rep3$ \\
  \hline
  $\rho_{\rep{r}}(\mathrm{S})$ & $1$     & $1$      & $1$        & $\frac13\begin{pmatrix}-1&2&-2\\2&-1&-2\\-2&-2&-1\end{pmatrix}$ \\[4mm]
  $\rho_{\rep{r}}(\mathrm{T})$ & $1$     & $\omega$ & $\omega^2$ & $\diag(1,\omega,\omega^2)$
\end{tabular}
\end{center}
The $A_4$ product rules are 
\begin{align*}
\rep1^a\otimes\rep1^b&=\rep1^c\ \ \text{with}\ \ c=a+b\!\!\!\mod3,\ \ 
\rep1^a\otimes\rep3=\rep3,\ \
\rep3\otimes\rep3 =\rep1\oplus\rep1'\oplus\rep1''\oplus\rep3\oplus\rep3,
\end{align*}
where $a,b,c=0,1,2$ correspond to the number of primes.

\subsection[\appendixname~\thesubsection]{ $T'$ }
\label{app:T'}

$T'$ (GAP Id~[24,3]) is the double cover of $A_4$ known also as binary tetrahedral group.
Its irreducible representations are $\rep{r}\in\{\rep1,\rep1',\rep1'',\rep2,\rep2',\rep2'',\rep3\}$.
This group can be generated by two generators $\mathrm{S}$ and $\mathrm{T}$ satisfying
$\mathrm{S}^4 = \mathrm{T}^3 = (\mathrm{ST})^3 = \Id$ and
$\mathrm{S}^2\mathrm{T}=\mathrm{T}\mathrm{S}^2$. This leads to the character table
\begin{center}
\begin{tabular}{c|ccccccc}
  class          & $1C_1$ & $1C_2$         & $6C_4$       & $4C_3$  & $4C_3'$ & $4C_6$ & $4C_6'$\\
  representative & $\Id$    & $\mathrm{S}^2$ & $\mathrm{S}$ & $\mathrm{T}$ & $\mathrm{T}^2$  & $\mathrm{S}^2\mathrm{T}$ & $\mathrm{S}^2\mathrm{T}^2$\\
  \hline
  $\rep1$   & $1$ & $1$  & $1$  & $1$        & $1$        & $1$        & $1$ \\
  $\rep1'$  & $1$ & $1$  & $1$  & $\omega$   & $\omega^2$ & $\omega$   & $\omega^2$\\
  $\rep1''$ & $1$ & $1$  & $1$  & $\omega^2$ & $\omega$   & $\omega^2$ & $\omega$ \\
  $\rep2$   & $2$ & $-2$ & $0$  & $-1$       & $-1$       & $1$        & $1$\\
  $\rep2'$  & $2$ & $-2$ & $0$  & $-\omega$  & $-\omega^2$& $\omega$   & $\omega^2$ \\
  $\rep2''$ & $2$ & $-2$ & $0$  & $-\omega^2$& $-\omega$  & $\omega^2$ & $\omega$\\
  $\rep3$   & $3$ & $3$  & $-1$ & $0$        & $0$        & $0$        & $0$
\end{tabular}
\end{center}
Note that the triplet representation is unfaithful; it yields only $A_4\cong T'/\Z2$, where
the normal \Z2 subgroup is generated by $\mathrm{S}^2$.
The representations can be expressed as
\begin{center}
\begin{tabular}{c|ccccccc}
  $\rep{r}$ & $\rep1$ & $\rep1'$ & $\rep1''$ & $\rep2$ & $\rep2'$ & $\rep2''$ & $\rep3$\\
  \hline
  $\rho_{\rep{r}}(\mathrm{S})$ & $1$ & $1$      & $1$        & $\Omega_\mathrm{S}$                               & $\Omega_\mathrm{S}$ & $\Omega_\mathrm{S}$           & $\rho_{\rep3}(\mathrm{S})$ \\
  $\rho_{\rep{r}}(\mathrm{T})$ & $1$ & $\omega$ & $\omega^2$ & $(\Omega_\mathrm{T}\widetilde\Omega_\mathrm{T})^*$& $\Omega_\mathrm{T}$ & $\widetilde\Omega_\mathrm{T}$ & $\rho_{\rep3}(\mathrm{T})$\\
\end{tabular}
\end{center}
where we defined the two-dimensional matrices
\begin{equation*}
  \Omega_\mathrm{S} = -\frac{\I}{\sqrt3}\begin{pmatrix}1&\sqrt2\\ \sqrt2&-1\end{pmatrix},\qquad
  \Omega_\mathrm{T} = \diag(1,\omega^2),\qquad
  \widetilde\Omega_\mathrm{T} = \diag(\omega,1)
\end{equation*}
and the three-dimensional representation is given (as in $A_4$) by
\begin{equation*}
\rho_{\rep3}(\mathrm S) =\frac13\begin{pmatrix}-1&2&-2\\2&-1&-2\\-2&-2&-1\end{pmatrix}
\qquad\text{and}\qquad
\rho_{\rep3}(\mathrm T) = \diag(1,\omega,\omega^2) \,.
\end{equation*}

Finally, the tensor products of the $T'$ irreducible representations are given by
\begin{align*}
\rep1^a\otimes\rep1^b&=\rep1^c,\quad \rep1^a\otimes\rep2^b=\rep2^c,\quad \rep2^a\otimes\rep2^b=\rep1^c\oplus\rep3
\quad \text{with}\quad c=a+b\mod3,\\
\rep1^a\otimes\rep3 &= \rep3,\quad
\rep2^a\otimes\rep3 = \rep2\oplus\rep2'\oplus\rep2''\quad\text{and}\quad
\rep3\otimes\rep3 =\rep1\oplus\rep1'\oplus\rep1''\oplus\rep3\oplus\rep3,
\end{align*}
where $a,b,c=0,1,2$ correspond to the number of primes. The Clebsch-Gordan coefficients can be found e.g.~in~\cite{Ishimori:2010au}.

\section[\appendixname~\thesection]{\boldmath Group theory elements of larger groups \unboldmath}
\label{app:deltas}

\subsection[\appendixname~\thesubsection]{$\Delta(54)$}
\label{app:delta54}

$\Delta(54)\cong(\Z3\x\Z3)\rtimes S_3\cong((\Z3\x\Z3)\rtimes\Z3)\rtimes\Z2$ (GAP Id~[54,8]) has 54 elements, which
can be generated by three generators $\mathrm{A},\mathrm{B},\mathrm{C}$ satisfying the presentation
$\mathrm{A}^3=\mathrm{B}^3=\mathrm{C}^2=(\mathrm{AB})^3=(\mathrm{AB}^2)^3=(\mathrm{AC})^2=(\mathrm{BC})^2=\Id$.
Its irreducible representations are two singlets, four doublets and two triplets plus their complex conjugates.
Together, they lead to the character table
\begin{center}
\begin{tabular}{c|cccccccccc}
  class      & $1C_1$ & $9C_2$       & $6C_3$       & $6C_3'$      & $6C_3''$        & $6C_3'''$        & $1C_3$           & $1C_3'$ 
                                                                                  & $9C_6$          & $9C_6'$\\
  repr.      & $\Id$    & $\mathrm{C}$ & $\mathrm{A}$ & $\mathrm{B}$ & $\mathrm{AB}$ & $\mathrm{AB}^2$ & $(\mathrm{ABC})^2$  & $(\mathrm{ACB})^2$
                                                                                  & $\mathrm{ABC}$  & $\mathrm{ACB}$\\
  \hline
  $\rep1$    & $1$ & $1$  & $1$  & $1$  & $1$  & $1$  & $1$         & $1$         & $1$         & $1$  \\
  $\rep1'$   & $1$ & $-1$ & $1$  & $1$  & $1$  & $1$  & $1$         & $1$         & $-1$        & $-1$  \\
  $\rep2_1$  & $2$ & $0$  & $2$  & $-1$ & $-1$ & $-1$ & $2$         & $2$         & $0$         & $0$  \\
  $\rep2_2$  & $2$ & $0$  & $-1$ & $2$  & $-1$ & $-1$ & $2$         & $2$         & $0$         & $0$  \\
  $\rep2_3$  & $2$ & $0$  & $-1$ & $-1$ & $-1$ & $2$  & $2$         & $2$         & $0$         & $0$  \\
  $\rep2_4$  & $2$ & $0$  & $-1$ & $-1$ & $2$  & $-1$ & $2$         & $2$         & $0$         & $0$  \\
  $\rep3_1$  & $3$ & $1$  & $0$  & $0$  & $0$  & $0$  & $3\omega$   & $3\omega^2$ & $\omega^2$  & $\omega$  \\
  $\rep3_2$  & $3$ & $-1$ & $0$  & $0$  & $0$  & $0$  & $3\omega$   & $3\omega^2$ & $-\omega^2$ & $-\omega$  \\
  $\crep3_1$ & $3$ & $1$  & $0$  & $0$  & $0$  & $0$  & $3\omega^2$ & $3\omega$   & $ \omega$   & $\omega^2$  \\
  $\crep3_2$ & $3$ & $-1$ & $0$  & $0$  & $0$  & $0$  & $3\omega^2$ & $3\omega$   & $-\omega$   & $-\omega^2$  
\end{tabular}
\end{center}
The doublets are unfaithful representations, which yield the quotient group $S_3\cong \Delta(54)/\Z3\x\Z3$,
where the normal subgroup \Z3\x\Z3 can be generated by $\mathrm{A}$ and $\mathrm{BAB}^2\mathrm{A}^2$.
In the irreducible representations, the $\Delta(54)$ generators can be expressed as
\begin{center}
\begin{tabular}{c|cccccccccc}
  $\rep{r}$ & $\rep1$ & $\rep1'$ & $\rep2_1$ & $\rep2_2$ & $\rep2_3$ & $\rep2_4$ & $\rep3_1$ & $\rep3_2$ & $\crep3_1$ & $\crep3_2$ \\
  \hline
  $\rho_{\rep{r}}(\mathrm{A})$ & $1$ & $1$  & $\Id_2$    & $\Omega_2$ & $\Omega_2$ & $\Omega_2$   & $\rho(A)$ & $\rho(A)$  & $\rho(A)$   & $\rho(A)$\\
  $\rho_{\rep{r}}(\mathrm{B})$ & $1$ & $1$  & $\Omega_2$ & $\Id_2$    & $\Omega_2$ & $\Omega_2^*$ & $\rho(B)$ & $\rho(B)$  & $\rho(B)^*$ & $\rho(B)^*$\\
  $\rho_{\rep{r}}(\mathrm{C})$ & $1$ & $-1$ & $S_2$      & $S_2$      & $S_2$      & $S_2$        & $\rho(C)$ & $-\rho(C)$ & $\rho(C)$   & $-\rho(C)$ \\
\end{tabular}
\end{center}
where the doublet representations are generated by
\begin{equation*}
\label{eq:mat2ofDelta54}
 \Id_2 = \begin{pmatrix}1&0\\0&1\end{pmatrix},\quad
 \Omega_2 = \begin{pmatrix}\omega^2&0\\0&\omega \end{pmatrix},\quad
 S_2 = \begin{pmatrix}0&1\\1&0\end{pmatrix},
\end{equation*}
and the triplets by
\begin{equation*}
\label{eq:mat3ofDelta54}
 \rho(A) = \begin{pmatrix}0&1&0\\0&0&1\\1&0&0\end{pmatrix},\quad
 \rho(B) = \begin{pmatrix}1&0&0\\0&\omega&0\\0&0&\omega^2\end{pmatrix},\quad
 \rho(C) = \begin{pmatrix}1&0&0\\0&0&1\\0&1&0\end{pmatrix}.
\end{equation*}
It is useful to list the nontrivial tensor products of $\Delta(54)$ irreducible representations:
\begin{align*}
\rep1'\otimes\rep1'&=\rep1,\ \ \rep1'\otimes\rep2_k=\rep2_k,\ \ \rep1'\otimes\rep3_1=\rep3_2,
\ \ \rep1'\otimes\rep3_2=\rep3_1,\ \ \rep1'\otimes\crep3_1=\crep3_2,\ \ \rep1'\otimes\crep3_1=\crep3_2,\\
\rep2_k\otimes\rep2_k&=\rep1\oplus\rep1'\oplus\rep2_k,\ \ \rep2_k\otimes\rep2_\ell =\rep2_m\oplus\rep2_n \quad \text{with}\ k\neq\ell\neq m\neq n, \ k,\ell,m,n=1,\ldots,4,\\
\rep2_k\otimes\rep3_\ell &=\rep3_1\oplus\rep3_2,\ \ \rep2_k\otimes\crep3_\ell=\crep3_1\oplus\crep3_2 \quad \text{for all}\ \ k=1,\ldots,4,\ \ell=1,2,\\
\rep3_\ell\otimes\rep3_\ell &= \crep3_1\oplus\crep3_1\oplus\crep3_2,\ \ \rep3_1\otimes\rep3_2 = \crep3_2\oplus\crep3_2\oplus\crep3_1,\ \
\rep3_1\otimes\crep3_1 = \rep1\oplus\rep2_1\oplus\rep2_2\oplus\rep2_3\oplus\rep2_4,\\
\rep3_1\otimes\crep3_2 &= \rep1'\oplus\rep2_1\oplus\rep2_2\oplus\rep2_3\oplus\rep2_4,\ \
\rep3_2\otimes\crep3_1 = \rep1'\oplus\rep2_1\oplus\rep2_2\oplus\rep2_3\oplus\rep2_4,\\
\rep3_2\otimes\crep3_2 &= \rep1\oplus\rep2_1\oplus\rep2_2\oplus\rep2_3\oplus\rep2_4,\ \
\crep3_\ell\otimes\crep3_\ell = \rep3_1\oplus\rep3_1\oplus\rep3_2,\ \
\crep3_1\otimes\crep3_2 = \rep3_2\oplus\rep3_2\oplus\rep3_1\,.
\end{align*}
The explicit Clebsch-Gordan coefficients can be found e.g.~in~\cite{Ishimori:2010au}.

\subsection[\appendixname~\thesubsection]{ $\Delta(27)$ }
\label{app:delta27}

The group $\Delta(27)\cong(\Z3\x\Z3)\rtimes\Z3$ (GAP Id~[27,3]) can be obtained from $\Delta(54)$, 
excluding the \Z2 generator $\mathrm C$. Hence, the generators $\mathrm A$ and $\mathrm B$ yielding
the 27 elements of the group are  constrained to fulfill only the subset of conditions 
$\mathrm A^3=\mathrm B^3=(\mathrm{AB})^3=(\mathrm{AB}^2)^3=\Id$. By excluding $\mathrm C$ in the
$\Delta(54)$ character table, we observe that the $\Delta(27)$ representations arise from the trivial
singlet, the doublets and combinations of triplets of $\Delta(54)$. One can show that they break into
nine singlets and two triplets, which describe the character table
\begin{center}
\resizebox{\textwidth}{!}{
\begin{tabular}{c|ccccccccccc}
  class      & $1C_1$ & $3C_3^a$  & $3C_3^b$  & $3C_3^c$   & $3C_3^d$  & $3C_3^e$  & $3C_3^f$  & $3C_3^g$ & $3C_3^h$ 
                                                                    & $3C_3^i$          & $1C_3$\\
  repr.      & $\Id$    & $\mathrm{A}$ & $\mathrm{B}$ & $\mathrm{A}^2$ & $\mathrm{B}^2$ & $\mathrm{AB}$ & $\mathrm{AB}^2$ & $(\mathrm{AB})^2$ & $\mathrm{BA}^2$
                                                                    & $\mathrm{A}^2\mathrm{B}^2\mathrm{AB}$ & $\mathrm{A}(\mathrm{AB})^2\mathrm{B}$\\
  \hline
  $\rep1_{0,0}$  & $1$ & $1$       & $1$       & $1$       & $1$       & $1$       & $1$       & $1$       & $1$        & $1$         & $1$    \\
  $\rep1_{0,1}$  & $1$ & $1$       & $\omega$  & $1$       & $\omega^2$& $\omega$  & $\omega^2$& $\omega^2$& $\omega$   & $1$         & $1$  \\
  $\rep1_{0,2}$  & $1$ & $1$       & $\omega^2$& $1$       & $\omega$  & $\omega^2$& $\omega$  & $\omega$  & $\omega^2$ & $1$         & $1$  \\
  $\rep1_{1,0}$  & $1$ & $\omega$  & $1$       & $\omega^2$& $1$       & $\omega$  & $\omega$  & $\omega^2$& $\omega^2$ & $1$         & $1$  \\
  $\rep1_{1,1}$  & $1$ & $\omega$  & $\omega$  & $\omega^2$& $\omega^2$& $\omega^2$& $1$       & $\omega$  & $1$        & $1$         & $1$  \\
  $\rep1_{1,2}$  & $1$ & $\omega$  & $\omega^2$& $\omega^2$& $\omega$  & $1$       & $\omega^2$& $1$       & $\omega$   & $1$         & $1$  \\
  $\rep1_{2,0}$  & $1$ & $\omega^2$& $1$       & $\omega$  & $1$       & $\omega^2$& $\omega^2$& $\omega$  & $\omega$   & $1$         & $1$  \\
  $\rep1_{2,1}$  & $1$ & $\omega^2$& $\omega^2$& $\omega$  & $\omega$  & $\omega$  & $1$       & $\omega^2$& $1$        & $1$         & $1$  \\
  $\rep1_{2,2}$  & $1$ & $\omega^2$& $\omega$  & $\omega$  & $\omega^2$& $1$       & $\omega$  & $1$       & $\omega^2$ & $1$         & $1$  \\
  $\rep3$        & $3$ & $0$       & $0$       & $0$       & $0$       & $0$       & $0$       & $0$       & $0$        & $3\omega$   & $3\omega^2$  \\
  $\crep3$       & $3$ & $0$       & $0$       & $0$       & $0$       & $0$       & $0$       & $0$       & $0$        & $3\omega^2$ & $3\omega$
\end{tabular}
}
\end{center}
Here we immediately see that the singlets $\rep1_{r,s}$, $r,s=0,1,2$, have the representations $\rho_{r,s}(\mathrm A) = \omega^{r}$
and $\rho_{r,s}(\mathrm B) = \omega^{s}$. Further, the triplet representations are given by $\rho_{\rep3}(\mathrm A) = \rho_{\crep3}(\mathrm A) = \rho(A)$
and $\rho_{\rep3}(\mathrm B) = \rho_{\crep3}(\mathrm B)^* = \rho(B)$, in terms of the $\Delta(54)$ matrices.

Finally, the tensor products of $\Delta(27)$ irreducible representations are given by
\begin{align*}
\rep1_{r,s}\otimes\rep1_{r',s'} &= \rep1_{r'',s''} \quad \text{with}\quad r''=r+r'\!\!\mod3,\ s''=s+s'\!\!\mod3,\\
\rep1_{r,s}\otimes\rep3 & =\rep3,\ \ \rep1_{r,s}\otimes\crep3  =\crep3 \quad \text{for all}\ \ r,s=0,1,2,\\
\rep3\otimes\rep3 &= \crep3\oplus\crep3\oplus\crep3,\ \ \crep3\otimes\crep3 = \rep3\oplus\rep3\oplus\rep3
\quad \text{and}\quad
\rep3\otimes\crep3 = \sum_{r,s}\rep1_{r,s}.
\end{align*}

\section[\appendixname~\thesection]{\boldmath $T'$ modular forms\unboldmath}
\label{app:TprimeForms}

The vector space of \SL{2}{\Z{}} modular forms of weight 1 associated with $T'\cong\Gamma_3'=\SL{2}{\Z{}}/\Gamma(3)$
can be spanned by~\cite{Liu:2019khw}
\begin{equation*}
\hat{e}_1(M) ~:=~ \frac{\eta^3(3\,M)}{\eta(M)} \qquad\mathrm{and}\qquad \hat{e}_2(M) ~:=~ \frac{\eta^3(M/3)}{\eta(M)}\;,
\end{equation*}
in terms of the Dedekind $\eta$-function of the modulus $M$. One can show that the combinations
\begin{equation*}
  Y^{(1)}(M) ~=~ \begin{pmatrix} \hat{Y}_1(M) \\ \hat{Y}_2(M) \end{pmatrix} ~:=~
\begin{pmatrix} -3\sqrt{2} & 0 \\ 3 & 1 \end{pmatrix} \begin{pmatrix} \hat{e}_1(M) \\ \hat{e}_2(M) \end{pmatrix}
\end{equation*}
transform under $\gamma\in\SL{2}{\Z{}}$ as
\begin{equation*}
  Y^{(1)}(M) ~\stackrel{\gamma}{\longrightarrow}~ (cM+d)\,\rho_{\rep2''}(\gamma)\, Y^{(1)}(M)\,,
\end{equation*}
i.e.\ building a $\rep2''$ representation $\rho_{\rep2''}$ of $\Gamma_3'\cong T'$, which is given in Appendix~\ref{app:T'}.
Higher-weight modular forms of $T'$ are derived from $Y^{(1)}(M)$ by the products of this basic vector-valued modular form,
such that $Y^{(n+m)}(M)=Y^{(n)}(M)\otimes Y^{(m)}(M)$. For instance, the modular forms of weight 2 are obtained from
$Y^{(2)}(M)=Y^{(1)}(M)\otimes Y^{(1)}(M)$, which build the $T'$ (and $A_4$) triplet\footnote{Other choices for the $T'$
Clebsch-Gordan coefficients lead to different but unimportant signs.}
$(\hat Y_2(M)^2,\sqrt2 \hat Y_1(M)\hat Y_2(M),\hat Y_1(M)^2)^\mathrm{T}=:(\hat X_1,\hat X_2,\hat X_3)^\mathrm{T}$.
The expected singlet from $\rep2''\otimes\rep2''=\rep1'\oplus\rep3$ vanishes, and we observe the known
relation $\hat X_2^2 - 2\hat X_1\hat X_3 = 0$, which can lead to interesting consequences~\cite{Chen:2024otk}.

\newpage
\begin{adjustwidth}{-\extralength}{0cm}

\reftitle{References}


\PublishersNote{}

\end{adjustwidth}
\end{document}